\documentclass[useAMS]{mn2e}
\usepackage{amssymb}
\usepackage{graphicx}

\title[Rotational suppression of the Tayler instability]
      {Rotational suppression of the Tayler instability
       in stellar radiation zones} 
\author[A. Bonanno and V. Urpin]
  {A.~Bonanno$^{1,2}$ and V.~Urpin$^{1,3,4}$ \\
      $^{1}$ INAF, Osservatorio Astrofisico di Catania,
      Via S.Sofia 78, 95123 Catania, Italy \\
      $^{2}$ INFN, Sezione di Catania, Via S.Sofia 72,
      95123 Catania, Italy \\
      $^{3}$ A.F.Ioffe Institute of Physics and Technology,
      194021 St. Petersburg, Russia \\
      $^{4)}$ Isaac Newton Institute of Chile, Branch in St. Petersburg,
           194021 St. Petersburg, Russia}
\date{today}

\pagerange{\pageref{firstpage}--\pageref{lastpage}} \pubyear{2002}

\def\LaTeX{L\kern-.36em\raise.3ex\hbox{a}\kern-.15em
    T\kern-.1667em\lower.7ex\hbox{E}\kern-.125emX}

\begin{document}

\label{firstpage}

\maketitle

\begin{abstract}
The study of the magnetic field in stellar radiation zones is an important topic in modern astrophysics 
because the magnetic field can play an important
role in several transport phenomena such as mixing and angular momentum transport.
We consider the influence of rotation on stability of a predominantly 
toroidal magnetic field in the radiation zone. We find that the effect of rotation 
on the stability depends on the magnetic configuration of the basic state. 
If the toroidal field  increases sufficiently rapidly with the spherical radius, the instability
cannot be suppressed entirely even by a very fast rotation although the strength of the 
instability can be significantly reduced. On the other hand, if the field increases slowly enough 
with the radius or decreases, the instability has a threshold and can be 
completely suppressed in rapidly rotating stars. We find that in the regions where the 
instability is entirely suppressed a particular type of magnetohydrodynamic 
waves may exist which are marginally stable.
\end{abstract}


\section{Introduction}  
In recent years, magnetic fields of various strength and topology 
have been detected in increasing number of stars (see, e.g., Donati 
et al. 2006). Despite the origin of stellar magnetic fields is
usually related to the turbulent dynamo action in convection zones, 
it is quite possible that magnetic fields exist  also in internal 
radiation zones as well. For instance, a uniform rotation of the 
solar core may be explained naturally by the presence of the magnetic 
field. A thin transition layer between the convection and radiation 
zones of the Sun, called the tachocline, cannot be understood within 
the frame of purely fluid-dynamical mechanisms and requires a large-scale 
magnetic field in the Sun's interior (Gough \& McIntyre 1998). The 
origin, topology, and strength of the magnetic field in internal radiation 
zones are the subject of debates for decades. Likely, dynamo cannot 
operate in stellar radiation zones. {A dynamo action requires 
sufficiently rapid flows with Re$_m \gg 1$ (Re$_m$ is the magnetic 
Reynolds number) to generate a global magnetic field but, according 
to the generally accepted point of view, such flows are not available 
in the internal radiation zones (see, e.g., Schwarzschild 1958). } 
{ Sometimes, the origin of the magnetic field in radiation zones is 
related to the dynamo mechanism proposed by Spruit (1999) but 
the existence of such mechanism has not be proven yet.} One might speculate that a fossil magnetic 
field could exist in radiation zones. Even a weak fossil field with 
non-vanishing poloidal component will quickly wrap up into a 
predominantly toroidal configuration, under the action of differential 
rotation. { Such configurations can be generated if Re$_m$ of 
differential rotation is greater than 1, or $| \nabla \Omega | > 
\eta_m / r^3$ where $\Omega$ and $\eta_m$ are the angular velocity and 
magnetic diffusivity, respectively. Estimating $|\nabla \Omega| \sim 
\Delta \Omega / r$ where $\Delta \Omega$ is a departure from the rigid 
rotation and assuming that the conductivity of plasma is $\sim 10^{16}$ 
s$^{-1}$, one can obtain that this condition is satisfied if $\Delta 
\Omega / \Omega > 10^{-18} \Omega_{sec}^{-1}$ where $\Omega_{sec}$ is the 
angular velocity in inverse seconds. Therefore, even very weak departures 
from the rigid rotation lead to a generation of the toroidal field. } The 
magnetic field with a predominant toroidal component is also typical 
for the liquid cores of neutron stars (Bonanno et al. 2005, 2006) 
where the field can be generated by the turbulent dynamo during the 
very early evolutionary stage. The toroidal field resulting from winding 
up or dynamo action cannot be arbitrarily strong because various magnetic 
instabilities will set in when it becomes strong enough. That is why 
much attention has been drawn to the instabilities that may affect 
the toroidal magnetic field in stably stratified radiation zones.

The magnetic field in a radiation zone can be subject to various 
instabilities such as the magnetic buoyancy (see, e.g., Gilman 1970, 
Acheson 1978) or magnetorotational instability (Velikhov 1959, Balbus 
1995). Likely, however, that the most efficient instabilities are 
caused by the electric currents maintaining the magnetic configuration 
(Spruit 1999). Such instabilities are well studied in cylindrical 
geometry in the context of laboratory fusion research (see, e.g., 
Freidberg 1970, Goedbloed \& Hagebeuk 1972). In astrophysical conditions, 
the instability caused by electric currents is studied mainly in 
cylindrical geometry as well (see, e.g., Tayler 1973). It 
turns out that the properties of instability depend on the ratio of 
the axial and toroidal fields and, even if this ratio is small, the 
axial field can alter the instability substantially (see, e.g., 
Knobloch 1992, Bonanno \& Urpin 2008a). The effect of an axial 
field on the Tayler instability of the toroidal field has been studied 
in detail in cylindrical geometry by Bonanno \& Urpin (2008b, 2011). 
The nonlinear evolution of the Tayler instability was considered by 
Bonanno et al. (2012) who argued that symmetry-breaking can give rise 
to a saturated state with non-zero helicity even if the initial state 
has zero helicity. A production of non-zero helicity is likely 
possibility for the onset of adynamo action.


Stability of the spherical magnetic configurations is much less studied 
because of mathematical problems. With numerical simulations Braithwaite 
\& Nordlund (2006) studied the stability of a random initial field and 
argued that this field relaxes on a stable mixed magnetic configuration 
with both poloidal and toroidal components. Stability of magnetic 
configurations with a predominantly toroidal field draws a particular 
attention. Numerical modeling by Braithwaite (2006) confirmed that the 
toroidal field with $B_{\varphi} \propto s$ or $\propto s^2$ ($s$ is the 
cylindrical radius) is unstable to the $m=1$ mode as it was predicted by 
Tayler (1973). It is widely believed that rotation and stratification 
can play an important role in stellar radiation zones providing a 
stabilizing influence on the Tayler instability (see, e.g., Pitts \& 
Tayler 1985, Spruit 1999). 
Recently, the effect of stratification and thermal conductivity on the 
Tayler instability of the toroidal field has been considered by Bonanno 
\& Urpin (2012). {The authors studied a linear stability and used a 
local approximation in latitude and global in the radial direction. They  
calculated the growth rate of instability and argued that the combined 
influence of gravity and thermal conductivity can never suppress the 
instability entirely. Stratification suppresses the instability at the pole
more efficiently than at the equator. The growth rate of instability can be 
essentially reduced by a stable stratification. A decrease of the 
growth rate caused by stratification is inversely proportional to the 
Brunt-V\"{a}is\"{a}l\"{a} frequency. Therefore, if gravity is strong,
the instability developes very slowly. A simple fitting expression has 
been obtained for the growth rate of instability in a stratified 
radiation zone. Bonanno \& Urpin (2012) argued that the instability of
modes with a large number of nodes in the radial direction is significantly 
more suppressed than the fundamental eigenmode. The reason of this is
qualitatively clear. As it was pointed out by Spruit (1999), the stabilizing
influence of stratification is less pronounced for perturbations with short
radial lengthscales. Therefore, it seems at the first glance that 
the instability should operate most efficiently on very short radial 
lengthscales. However, this conclusion is incorrect because a destabilizing 
effect of electric currents (that is the reason of the Tayler instability) 
also decreases if the radial lengthscale decreases. This occurs because the 
instability becomes almost two-dimensional if the radial lengthscale goes 
to 0 but, as it was shown by Kitchatinov \& R\"udiger (2008), the  Tayler 
instability does not exist in 2D. Therefore, the instability cannot arise 
for perturbations with very short radial lengthscales despite the 
stabilizing influence of stratification is minimal for them.} 

Rotation can also suppress the Tayler instability and stabilize the 
magnetic configurations. The effect of rotation has been considered by
a number of authors. Spruit (1999) found that the growth rate of the Tayler 
instability near the rotation axis should be of the order of $\sim \omega_A 
(\omega_A /\Omega)$ if $\Omega \gg \omega_A$ in a particular case of the
magnetic field dependent on $s$ alone, where $\omega_A$ 
and $\Omega$ are the Alfven frequency and angular velocity of the star, 
respectively. In numerical modelling by Braithwaite (2006), the Tayler 
instability was suppressed if $\Omega$ is above a certain value of the 
order of $\omega_A$. Above this value a distinct oscillatory behaviour 
sets in with marginal stability. Stability of the toroidal field in 
rotating radiation zones has been considered also by Kitchatinov (2008) 
and Kitchatinov \& R\"udiger (2008) who clarified that the Tayler's
instability recovers only in 3D. Radial displacements are essential
for this instability. It does not exist in the 2D case of strictly
horizontal (perpendicular to gravity) disturbances and only stable 
modes exist in this case. These authors argued that the magnetic 
instability is characterized by the threshold field strength. The
instability arises if the magnetic field is stronger than this threshold
but it does not occur if the field is weaker than the threshold. 
Estimating this threshold in the solar radiation zone, Kitchatinov \& 
R\"udiger (2008) impose the upper limit on the magnetic field $\approx 
600$ G. The stability of the toroidal field in a rotating radiation zone has 
been studied also by Zahn et al. (2007) in the particular case $B_{\varphi} 
\propto s$. The particular type of oscillatory modes found by these authors 
is relevant to rotation and is stable in the non-dissipative limit. 
However, instability of the considered modes can occur in a form of an 
oscillatory diffusive instability if dissipation is provided by radiative 
or Ohmic diffusion. Unfortunately, the authors of these studies did not 
compare the results of their calculations and the reason of such 
qualitatively different behaviours is unclear. 

In this paper, we consider in detail the effect of rotational suppression 
of the Tayler instability in the case of a predominantly toroidal field. 
We show that rotation can influence the instability in different ways 
depending on the baseground magnetic configuration. The paper is organized 
as follows. The basic equations and mathematical formulation of the 
problem are presented in Sec.2. This is followed by results of numerical 
calculations of the growth rate and frequency of unstable modes in Sec.3. 
The paper closes with a summary of the main results and some remarks in 
Sec.4.

\section{Basic equations}

We assume that the magnetic field in a radiation zone is subthermal and 
the magnetic pressure is smaller than the gas pressure. In this case,
the Boussinesq approximation is applied for a consideration of low-frequency 
modes. The ideal MHD equations read in this approximation   
\begin{eqnarray}
\frac{\partial \vec{v}}{\partial t} + (\vec{v} \cdot \nabla) \vec{v} = 
- \frac{\nabla p}{\rho} + \vec{g} 
+ \frac{1}{4 \pi \rho} (\nabla \times \vec{B}) \times \vec{B}, 
\end{eqnarray}
\begin{equation}
\frac{\partial \vec{B}}{\partial t} - \nabla \times (\vec{v} \times \vec{B}) 
= 0,
\end{equation}
\begin{equation}
\nabla \cdot \vec{v} = 0, \;\;\; \nabla \cdot \vec{B} = 0, 
\end{equation}
where $\vec{g}$ is gravity, $p$ and $\rho$ are the gas pressure and density, 
respectively. The equation of thermal balance reads in the Boussinesq 
approximation
\begin{equation}
\frac{\partial T}{\partial t} + \vec{v} \cdot (\nabla T - \nabla_{ad}T) =
0,
\end{equation} 
where $\nabla_{ad} T$ is the adiabatic temperature gradient.

Consider the stability of an axisymmetric toroidal magnetic field using 
spherical coordinates ($r$, $\theta$, $\varphi$). We assume that the 
radiation zone rotates with the angular velocity $\vec{\Omega}$=const and 
that the toroidal field depends on $r$ and $\theta$, $B_{\varphi}= 
B_{\varphi}(r, \theta)$. In the unperturbed state, the radiation zone is 
assumed to be in hydrostatic equilibrium, then
\begin{equation}
\frac{\nabla p}{\rho} = \vec{g} + \frac{1}{4 \pi \rho} 
(\nabla \times \vec{B}) \times \vec{B} + \vec{e}_s \;\Omega^2 \;r \sin \theta,
\end{equation}
where $\vec{e}_s$ is the unit vector in the cylindrical radial direction.
The rotational energy is assumed to be much smaller than the gravitational
one, $g \gg r \Omega^2$. Since the magnetic energy is subthermal, $\vec{g}$ 
is approximately radial in our basic state. Therefore, longitudinal variations
of the unperturbed density and pressure are small.

We consider a linear stability. Small perturbations will be indicated by 
subscript 1, while unperturbed quantities will have no subscript. Linearizing 
Eqs.(1)-(4), we take into account that small perturbations of the 
density and temperature in the Boussinesq approximation are related by
$\rho_1/\rho = -\beta (T_1/T)$ where $\beta$ is the thermal expansion 
coefficient. We use a local approximation in the $\theta$-direction and 
assume that small perturbations depend on $\theta$ as $\propto \exp( - i 
l \theta)$, where $l \gg 1$ is the polar wavenumber. Since the basic state 
is stationary and axisymmetric, the dependence of perturbations on $t$ and 
$\varphi$ can be taken in the exponential form as well. Then, perturbations 
are proportional to $\exp{(\sigma t - i l \theta - i m \varphi)}$ where $m$ 
is the azimuthal wavenumber and $\sigma$ is the growth rate. The dependence 
of perturbations on $r$ should be determined from Eqs.(1)-(4).

Since the influence of gravity on instability has been studied in detail 
by Bonanno \& Urpin (2012), we concentrate in this paper mainly on the 
effect of rotation. We consider a simplified problem assuming that 
stratification is neutral and $\nabla T = \nabla_{ad}T$. In this case, 
the stabilizing effect of stratification is excluded and we can study 
only the influence of rotation alone. The combined influence of
rotation and stratification will be considered elsewhere. For the sake 
of simplicity, we also assume that the unperturbed density is approximately 
homogeneous in the radiation zone. 

Eliminating all variables in favor of $v_{1r}$, we obtain with the accuracy 
in terms of the lowest order in $(k_{\theta} r)^{-1}$ 
\begin{eqnarray}
(\sigma_0^2 + \omega_A^2 + D \Omega_i^2) \; v_{1r}'' +
\left( \frac{4}{r} \sigma_0^2 + \frac{2}{H} \omega_A^2  \right) v_{1r}' +
\\ \nonumber 
\left[ \frac{2}{r^2} \sigma_0^2
- k_{\perp}^2 (\sigma_0^2 +  \omega_A^2 ) - D \Omega_e^2 k_{\theta}^2 +
\frac{2}{r} \omega_A^2 \left( \frac{1}{H} 
\frac{k_{\perp}^2}{k_{\varphi}^2} - 
\right.
\right.
\\ \nonumber
\left.
\left.
\frac{2}{r} 
\frac{k_{\theta}^2}{k_{\varphi}^2} D \right)
- i \sigma_0 \Omega_e \left( \frac{k_{\varphi}}{r} +
4 D \frac{k_{\theta}^2}{r k_{\varphi}} \frac{\omega_A^2}{\sigma_0^2}
\right) \right] v_{1r} = 0,
\end{eqnarray}
where the prime denotes a derivative with respect to $r$ and
\begin{eqnarray}
\sigma_0 = \sigma - i m \Omega, \;\;\; 
\omega_A^2 = \frac{k_{\varphi}^2 B_{\varphi}^2}{4 \pi \rho}, \;\;\;
D = \frac{\sigma_0^2}{\sigma_0^2 + \omega_A^2}, 
\\ \nonumber
\Omega_i=2 \Omega \cos \theta,\;\;\; \Omega_e=2 \Omega \sin \theta,\;\;\; 
k_{\perp}^2 = k_{\theta}^2 + k_{\varphi}^2, \;\;\;
\\ \nonumber
\frac{1}{H} = \frac{\partial}{\partial r} \ln (r B_{\varphi}). 
\end{eqnarray}
The polar and azimuthal wavevectors are $k_{\theta}=l/r$ and 
$k_{\varphi}=m/r \sin \theta$, respectively.  

This equation with the corresponding boundary conditions describes the 
stability problem as a non-linear eigenvalue problem. Fortunately, the 
main qualitative features of this problem are not sensitive to the choice 
of boundary conditions. That is why we choose the simplest boundary 
conditions and assume that $v_{1r}= 0$ at the inner and outer boundaries,
$r=R_i$ and $r=R$, respectively.

\section{Numerical results}

We assume that the radiation zone is located at $R_i \leq r \leq R$ or,
introducing the dimensionless radius $x= r/R$, at $x_i \leq x \leq 1$
where $x_i= R_i/R$. We choose the internal radius of the radiation zone, 
$x_i$, to be small but finite for computational convenience. 
{In particular
$x_i=0.1$ in this investigation.
Our results 
do not depend qualitatively on the precise value of $x_i$, 
as we have explicitly verified by performing 
calculations with with  $x_i$ in the range $0.1\leq x_i \leq 0.4$
(less than $1\%$ difference in the growth rate 
for $x_i=0.1$ and  $x_i=0.3$ for $\alpha=2$ for instance).}

We represent the toroidal field as
\begin{equation}
B_{\varphi} = B_0 (x/x_i)^{\alpha} \sin \theta,
\end{equation}
where $B_0$ is the field strength at $x=x_i$ at the equator. The 
dependence of $B_{\varphi}$ on $r$ is uncertain in the radiation zone. 
Therefore, we consider different possibilities, varying the parameter
$\alpha$. The radial dependence in Eq.(8) is the simplest one and 
convenient from the computational point of view. A power law
dependence on the cylindrical radius was also used, for example, in 
the pioneering paper by Tayler (1973) and, therefore, it is more 
convenient to compare the results for spherical and cylindrical
geometries, choosing the dependence (8). Eq.(8) can mimic various 
physical situations. The case $\alpha > 0$, for instance, can be a
representative of the radiation zone in a star with the outer
convective zone. On the contrary, the case $\alpha < 0$ can mimic 
a radiation zone with the magnetic field generated in the inner 
convective core. Certainly, the magnetic field in a radiation zone
can have a more complicated dependence than Eq.(8) but the main 
qualitative features can be understood from this simple model.

It is convenient to introduce the dimensionless quantities 
\begin{equation}
\Gamma = \frac{\sigma_0}{\omega_{A0}}, \;\;\; 
\eta = \frac{2 \Omega}{\omega_{A0}}, 
\end{equation} 
where $\omega_{A0} = B_{0} /R \sqrt{4 \pi \rho}$. We will represent the 
results in terms of $\Gamma$ and $\eta$.

In Fig.~1, we plot the growth rate and frequency of unstable modes
as functions of $\eta$ for different $\theta$ and $\alpha = 3$. Such 
distribution of $B_{\varphi}$ can mimic, for example, the radiation zone 
of a star with a convective envelope. In this case, the bottom of a 
convection zone is likely the location of the toroidal field generated 
by a dynamo action. However, this case can be also the representative 
of a star with relic magnetic fields. Details of the formation of such 
fields are very uncertain but it is quite possible that differential 
rotation is stronger in the outer layers at the early stage of stellar 
evolution. Then, the toroidal field generated by differential rotation 
should be stronger in the outer layers as well. 

Fig.~1 shows that the instability is most efficient at the equator and 
does not occur around the pole. There always exist a range of $\theta$
around the pole where the instability does not occur and this range 
depends on $\alpha$. Note that the instability exhibits a similar 
behaviour also in the non-rotating case (see Bonanno \& Urpin 2012). 
Our result is at variance with a widely accepted opinion 
that toroidal magnetic configurations are always unstable at the axis 
(see, e.g., Spruit 1999). This opinion is based on similarity of 
the spherical magnetic configuration near the axis and the axisymmetric 
cylindrical configuration. However, this analogy is generally incorrect 
because, in spherical geometry, the toroidal field near the axis depends 
also on distance along the magnetic axis and not only on the distance 
from it. This dependence plays the crucial role in stability. At small 
$\eta$, the maximum growth rate is of the order of  $\omega_{A0}$ and is 
reached at the equator. However, the growth rate clearly shows some 
suppression for a faster rotation. Suppression becomes important already at 
relatively small values of $\eta \sim 1-2$. We can distinguish two 
substantially different regimes of suppression.
If $\theta$ is greater than some characteristic value, $\theta_0$, 
the growth rate decreases with an increase of $\eta$ approximately as 
$1/ \eta$ and it does not vanish even at very large $\eta$. It turns 
out, therefore, that rotation can never suppress the Tayler instability 
in the region around the equator, where $\theta > \theta_0$, but can 
only decrease its growth rate. Note that the Tayler's modes are complex 
in this region in contrast to the non-rotating case. The frequency (= Im
$\Gamma$) is basically comparable to the growth rate and also decreases 
if rotation becomes faster. Note also that  similar dependence of 
the growth rate on $\eta$ was obtained by Spruit (1999) who considered 
the instability in cylindrical geometry with the toroidal field dependent 
on the cylindrical radius alone, $B_{\varphi} = B_{\varphi}(s)$ where $s$ 
is the cylindrical radius. The exact value of $\theta_0$ is rather 
difficult to calculate because of computational problems but, in the 
case $\alpha = 3$, it is equal approximately to $26^{\circ}$. For $\theta 
< \theta_0$, the effect of 
rotation is substantially stronger. In contrast to the instability near 
the equator, the instability at $\theta < \theta_0$ is characterized by 
the threshold, $\eta_{cr}$. The instability occurs if $\eta < 
\eta_{cr}$ (Re $\Gamma >0$ in this region) but it does not occur if 
$\eta > \eta_{cr}$ since $\Gamma$ is imaginary at such $\eta$. The 
threshold, $\eta_{cr}$, depends on $\theta$ and is $\sim 1$ if $\theta 
= 24^{\circ}$. The threshold value $\eta_{cr}$ is lower for smaller 
$\theta$. Therefore, even a relatively slow rotation with $\eta < 1$ 
can suppress the instability entirely in some region around the pole. 
Note that the Tayler modes become oscillatory beyond the threshold 
(at $\eta > \eta_{cr}$): their growth rate is equal to zero but the 
frequency is always non-vanishing. In the region around the pole 
(at $\theta < \theta_{cr}$), the instability does not arise even if
rotation is very slow ($\eta \ll 1$). The critical value of the polar
angle, $\theta_{cr}$, that restricts the stable region for a 
non-rotating star, is $\sim 21^{\circ}$.

\begin{figure}
\includegraphics[width=9cm]{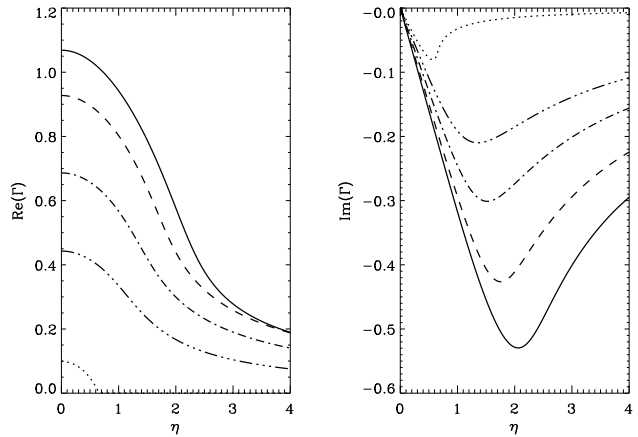}
\caption{The growth rate (left panel) and frequency (right panel) 
of the Tayler's modes as functions of the rotational parameter $\eta$
for $\alpha = 3$ and for $\theta = 90^{\circ}$ (solid), $63^{\circ}$ 
(dashed), $46^{\circ}$ (dash-and-dotted), $34^{\circ}$ (dash-dot-dotted), 
and $24^{\circ}$ (dotted). The longitudinal wavenumber is $l = 10$.}
\end{figure}

In Fig.~2, we plot the growth rate and frequency for the case $\alpha
= 2$. Qualitatively, the behaviour is same but suppression turns out
to be more essential. Again, the instability is most efficient at the 
equator and does not occur around the pole. This is qualitatively 
clear because, on the equator, with a neutral entropy gradient, we 
have an equivalent situation to the cylindrical geometry. The growth 
rate is maximal 
for small $\eta$ and is $\approx \omega_{A0}$ at the equator that is a 
bit lower than in the case $\alpha = 3$. Rotational suppression 
becomes important already at $\eta \sim 1$ and can operate in two 
different regimes, depending on $\theta$. The characteristic value 
$\theta_0$ that distinguishes these two regimes is $\approx 37^{\circ}$
in this case. Unfortunately, we cannot calculate the growth rate when
both $\theta$ and $\eta$ are close to their critical values, $\theta_{0}$ 
and $\eta_{cr}$, respectively. This is caused by computational problems.  
At $\theta > \theta_0$, the growth rate decreases with 
an increase of $\eta$ approximately as $1/ \eta$ and it does not 
vanish at large $\eta$. Rotation can never suppress the instability 
in this region of $\theta$ but can only reduce its growth rate. Like 
the case $\alpha=3$, unstable modes are oscillatory in this region 
with the frequency being comparable to the growth rate. Closer to the 
pole, at $\theta < \theta_0$, the instability is characterized by the 
threshold, $\eta_{cr}$: the instability can occur if $\eta < \eta_{cr}$ 
but it is suppressed if $\eta > \eta_{cr}$. For example, at $\theta = 
37^{\circ}$, the critical value $\eta_{cr}$ is $\approx 3.5$ but
it is lower for smaller $\theta$. In the region near the pole, 
$\theta < \theta_{cr} = 24^{\circ}$, the instability is completely 
suppressed and does not occur at any $\eta$. However, there exist 
stable oscillatory modes with a low frequency. 

\begin{figure}
\includegraphics[width=9cm]{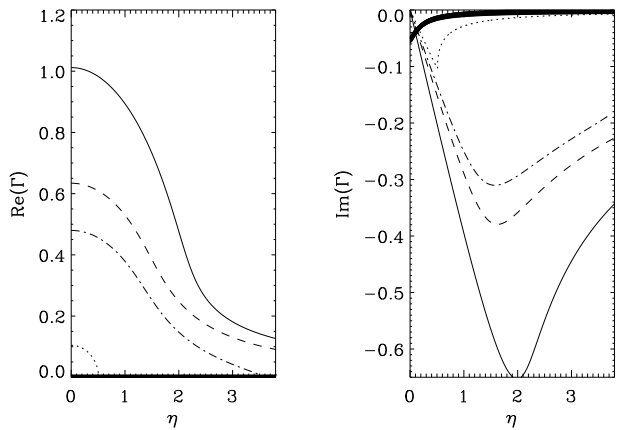}
\caption{The growth rate (left panel) and frequency (right panel) 
of the Tayler's modes as functions of the rotational parameter $\eta$
for $\alpha = 2$ and for $\theta = 90^{\circ}$ (solid), $45^{\circ}$ 
(dashed), $37^{\circ}$ (dash-and-dotted), $26^{\circ}$ (dotted), and 
$23^{\circ}$ (thick solid). The longitudinal wavenumber is $l = 10$.}
\end{figure}

Fig.~3 plots the growth rate and frequency for the toroidal 
field with $\alpha = 1$. Like the previous cases, the instability is 
most efficient at the equator and its growth rate decreases if 
$\theta$ decreases. Generally, the instability at $\alpha = 1$ is 
suppressed stronger than in the cases $\alpha = 3$ and $\alpha = 2$. 
However, there is a qualitative difference in comparison to these 
cases: there is no regime near the equator in which the growth rate 
is non-vanishing at large $\eta$ or, in other words, $\theta_0 = 
\pi/2$. It turns out that such regime of instability can occur only 
at a relatively large $\alpha$. The instability at $\alpha = 1$ is 
characterized by the threshold, $\eta_{cr}$, even in the region near 
the equator. The threshold is lower than for $\alpha = 2$ and is 
equal to $\approx 2$ at the equator. As usual, the threshold is 
lower for smaller $\theta$. The field near the magnetic axis turns 
out to be stable. The instability does not occur at any $\eta$ in 
the region with $\theta < \theta_{cr} \approx 31^{\circ}$. 

The condition of stability $\eta > \eta_{cr}$ can easily be 
reformulated in terms of the angular velocity and magnetic 
field. Rotation suppresses completely the Tayler instability if the 
star rotates with the angular velocity  
\begin{equation}
\Omega > \frac{\eta_{cr}}{2} \cdot \frac{B_0}{R \sqrt{4 \pi \rho}}.
\end{equation} 
We can also rewrite this inequality as the condition for the
magnetic field,
\begin{equation}
B_0 < \frac{2}{\eta_{cr}} \Omega R \sqrt{4 \pi \rho}.
\end{equation}

The Tayler modes become oscillatory (Im $\Gamma=0$) if condition (10) or
(11) are satisfied. Note that such behaviour was also seen in numerical
modelling of the Tayler instability by Braithwaite (2006). The frequency 
of marginally stable waves is of the order of $\omega_{A0} (\omega_{A0} / 
\Omega)$ at $\eta > 1$ and it decreases $\propto 1/ \eta$ at large 
$\eta$. This new type of magnetohydrodynamic waves is determined by the 
Coriolis and Lorentz force and can exist at those values of $\eta$ 
which suppress the Tayler instability. The dispersion relation for 
these waves can easily be obtained from Eq.(6) in short wavelength 
approximation. Consider perturbations with a very short radial 
wavelength for which one can use a local approximation in the radial 
direction, such as $v_{1r} \propto \exp( -ik_{r} r)$, where $k_{r}$ is 
the radial wavevector. If $k_{r} \gg \max(k_{\theta}, k_{\varphi})$, 
then Eq.(6) yields with the accuracy in terms of the lowest order in 
$(k_{r} r)^{-1}$ the following dispersion equation
\begin{equation}
\sigma_0^2 + \omega_A^2 + D \Omega_i^2 = 0, 
\end{equation}
or
\begin{equation}
\sigma_1^4 + \sigma_1^2 (2 \omega_A^2 + \Omega_i^2) + \omega_A^4 
= 0.
\end{equation}
The solution of this equation is 
\begin{equation}
\sigma_1^2 = - \frac{1}{2} (\Omega_i^2 + 2 \omega_{A}^2 ) \pm \frac{1}{2}
\Omega_i^2 \sqrt{1 + \frac{4 \omega_A^2}{\Omega_i^2}}.
\end{equation}
If $\Omega_i \gg \omega_A$ then we can expand a square root in a power series
of $(\omega_A/\Omega_i)^2$. Then, choosing the upper sign in Eq.(14), we 
obtain with accuracy in the lowest order in $(\omega_A/\Omega_i)^2$
\begin{equation}
\sigma_1^2 \approx - \omega_A^2 (\omega_A/\Omega_i)^2.
\end{equation} 
This dispersion relation describes a new type of oscillatory modes that
can exist in rapidly rotating stars. These modes can be called the
``magneto-inertial'' waves because they can exist only in a magnetized 
and rotating plasma. In the considered case, these waves are stable
but, likely, they can be unstable under certain conditions. We
consider the magneto-inertial waves in more detail elsewhere.

\begin{figure}
\includegraphics[width=9cm]{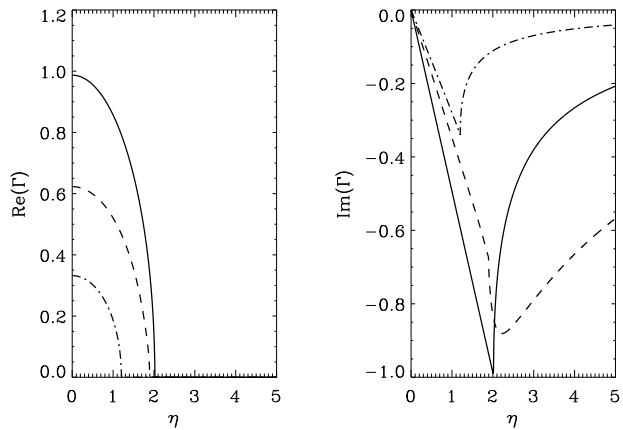}
\caption{The same as in Fig.1 but for $\alpha = 1$. The curves correspond
to $\theta = 90^{\circ}$ (solid), $45^{\circ}$ 
(dashed), and $37^{\circ}$ (dash-and-dotted).  }
\end{figure}

Fig.~4 shows the growth rate and frequency for $\alpha = - 0.4$. 
The instability is very much suppressed by rotation in this case.
It can occur only in a narrow region around the equator, 
$90^{\circ} \geq \theta > \theta_{cr} \approx 65^{\circ}$, and does not 
occur in the extended region around the rotation axis, $\theta < 
\theta_{cr} \approx 65^{\circ}$. The threshould value of $\eta$ at the
equator is $\approx 2$. Therefore, the Tayler instability is entirely 
suppressed everywhere in the radiation zone with $\alpha = -0.4$ if 
$\eta > 2$, or $\Omega > \omega_{A0}$. However, stable oscillating 
modes can exist even at much higher $\eta$.

Magnetic configurations with a rapidly decreasing toroidal field are 
stable in a cylindrical geometry. For example, the field can be 
unstable to non-axisymmetric perturbations only if $d \ln B_{\varphi}(s) 
/ d \ln s > - 1/2$ (Tayler 1973a). Our calculations do not show the 
presence of instability in the spherical geometry if $\alpha < - 1/2$. 
 
\begin{figure}
\includegraphics[width=9cm]{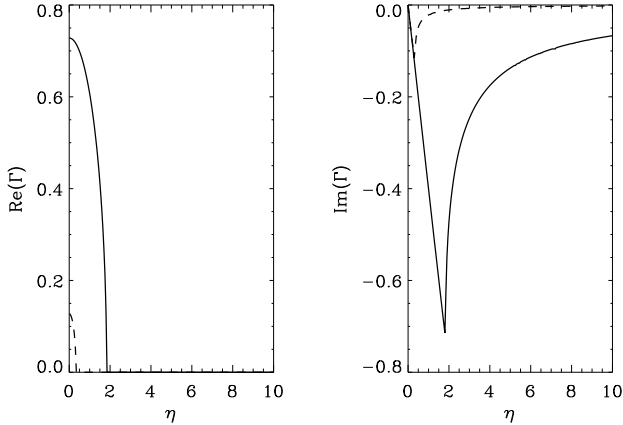}
\caption{The same as in Fig.2 but for $\alpha = - 0.4$ and $\theta = 
90^{\circ}$ (solid), and $69^{\circ}$ (dashed).}
\end{figure}

Fig.~5 plots two examples of eigenfunctions for $\alpha = 2$ and
$\eta = 2$. The eigenfunction at $\theta = 90^{\circ}$ corresponds
to the unstable mode that cannot be suppressed by rotation. Motions
in this mode take place basically in the outer part of the radiation 
zone. Note that, at larger $\eta$, this mode (corresponding to 
$\theta = 90^{\circ}$) turns out to be localized in a more narrow 
region near the outer boundary. On the contrary, the
eigenfunction at $\theta = 26^{\circ}$ corresponds to a stable 
oscillating mode that exists beyond the instability threshold, 
$\eta > \eta_{cr}$. In this case, the mode tends to be located near
the inner boundary. This type of modes also becomes sharper with 
increasing $\eta$. Note that this seems to be a rather general 
rule (at least, we did not find an exception): modes tend to be 
localized in the outer part of the radiation zones for $\theta > 
\theta_0$ and, on the contrary, they are localized in the inner part
for $\theta < \theta_0$. Therefore, for example, in the cases 
$\alpha = 1$ or $-0.4$, where the region with $\theta > \theta_0$ 
is absent, both stable and unstable modes are localized closer to 
the inner boundary.


\begin{figure}
\includegraphics[width=9cm]{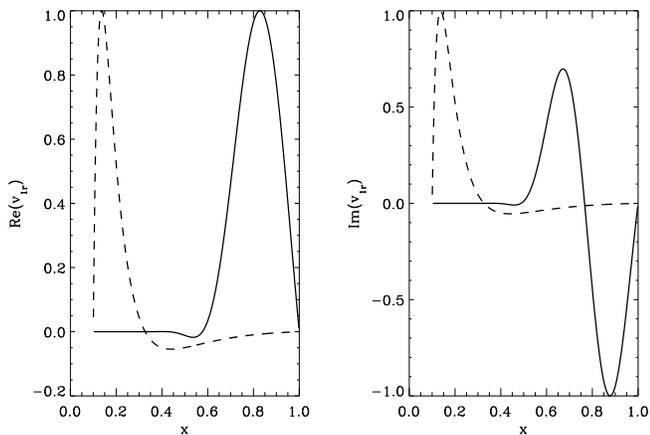}
\caption{The eigenfunctions $v_{1r}$ $\alpha = 2$, $\eta =2$, and 
$\theta = 90^{\circ}$ (solid) and $26^{\circ}$ (dashed).}
\end{figure}

\section{Discussion}

We have considered the stability of the toroidal field in rotating 
stellar radiation zones. Stability properties of the spherical magnetic 
configurations are generally qualitatively different from those of the 
cylindrical configurations. For instance, our calculations show that the 
instability is most efficient near the equator but the toroidal field 
can be stable near the symmetry axis. This is in the contrast to a widely 
accepted opinion based on the analysis of cylindrical configurations
(see, e.g., Spruit 1999, Zhang et al. 2003) that the toroidal field is
always unstable near the axis in the spherical geometry. { However, 
a direct analogy between stability of an infinitely long cylinder with 
the azimuthal field and the toroidal field in stellar radiation zones 
is generally misleading. In order to apply the cylindrical geometry 
near the axis of a spherical star, one needs to make sure that i) the 
length scale of perturbations in the axial direction is small compared 
to the radius and ii) the radial (cylindrical) length scale of 
perturbations is small compared to the radius. It is clear that 
perturbations should be affected by the spherical geometry if their 
radial length scale is comparable to the stellar radius. Therefore, 
the behaviour of such modes with a large radial length scale may be
different in the spherical and cylindrical geometries. As far as 
perturbations with a small radial length scale are concerned, they 
are alway stable (see Eq.(14)-(15)). Therefore, the analogy with an 
infinitely long cylinder is generally incorrect and the Tayler instability 
can be suppressed near the magnetic axis in the spherical geometry.
It seems that a dependence on the axial coordinate always makes 
perturbations more stable. }


Rotation provides an additional stabilizing influence on the Tayler 
instability. The effect of rotation is characterized by the parameter 
$\eta = 2 \Omega / \omega_{A0}$ that can be large in radiation zones 
{\bf ($\sim 10^2-10^3$ if $B \sim 10^4$ G)}. A reduction of the growth 
rate becomes significant already at a relatively low angular velocity, 
$\Omega \sim \omega_{A0}$, that corresponds to $\eta \sim 1$. It turns out 
that the effect of rotation can be twofold and depends crucially on the radial 
dependence of the toroidal field. If the toroidal field increases with the 
spherical radius sufficiently rapidly ($\alpha = 2-3$) then rotation cannot 
suppress completely the Tayler instability in the region around the equator,
even if $\eta$ is very large. This region is more extended for larger 
$\alpha$ but shrinks around the equator and disappear for smaller $\alpha$. 
The growth rate of instability in this region is non-vanishing for any 
angular velocity but it can be substantially reduced at large $\eta$. At 
large $\eta$, the growth rate descreases $\propto 1/\eta$ and is 
approximately equal to $\omega_{A0} (\omega_{A0} / \Omega)$. This expression 
was also obtained by Spruit (1999) for the growth rate near the rotation 
axis in a particular case $B_{\varphi} = B_{\varphi}(s)$. Note, however, 
that the Alfven timescale, $\omega_{A0}^{-1}$, is typically short ($\sim 
1-3$ yrs if $B \sim 10^3$ G) compared to the stellar life-time, therefore 
even a suppressed instability with a reduced growth rate can be important 
in radiation zones. 

Such region near the equator exists only if $\alpha$ is sufficiently 
large ($\gtrsim 1.5-2$) but it shrinks and disappears if $\alpha$ decreases.
At smaller $\theta$, the region is located where the rotational suppression 
is qualitatively different and where the instability is determined by the 
threshold value of $\eta$. The Tayler instability turns out to be suppressed 
completely in this region if $\eta > \eta_{cr} \gtrsim 1-2$. Therefore, 
modes are stable everywhere in this region 
for such rapidly rotating stars. The threshold is not very high and 
corresponds approximately to $\Omega \sim \omega_{A0}$. Higher eigenmodes 
are suppressed stronger than the fundamental one and perturbations with 
a short radial wavelength are always stable. 
Since the instability is suppressed at $\eta > \eta_{cr}$ but not
suppressed at $\eta < \eta_{cr}$, this implies that the magnetic field 
should satisfy the condition 
\begin{equation}
B_0 \gtrsim \frac{2}{\eta_{cr}} \Omega R \sqrt{ 4 \pi \rho} 
\end{equation}
in order the instability could occur. Estimating $\Omega R \sim 2 \times 
10^5$ cm/s and $\rho \sim 0.1$ g/cm$^3$, we obtain that instability can 
arise in the radiation zone of the Sun if $B_0 \gtrsim 10^5$ G. This 
estimate of the critical field is more than two orders of magnitude 
higher than that obtained by Kichatinov \& R\"udiger (2008).

{ The hydrodynamic motions generated by instability can be important
for the transport processes in radiation zones. Motions in the unstable 
modes depend generally on the polar and azimuthal wavenumbers as well as
on the parameter $\eta$. In real conditions, $\eta$ is likely large 
($\sim 10^3 - 10^4$) and the radial length scale of eigenmodes is rather 
short. If it is shorter than $r/l$ or $l/m$, then one can easily estimate 
from the continuity equation that the radial component of velocity is
small compared to the polar and azimuthal ones. However, the character
of motions can be different at the non-linear stage when the interaction 
between modes becomes essential.}

It turns out that a new type of magneto-inertial waves might exist in
the radiation zones of rapidly rotating stars in the regions where 
$\eta > \eta_{cr}$ and the Tayler instability is suppressed. These waves 
are marginally stable and their dispersion relation is given by Eq.(15) 
at large $\eta$. The frequency of this oscillation decreases with an 
increase of $\Omega$. These waves are stable in our simplified model 
but they can be unstable in more realistic conditions (the presence of 
a poloidal field, differential rotation, etc.). The magneto-inertial 
waves can play an important in various processes in the radiation zone 
(mixing, transport of the angular momentum, formation of the tachocline, 
etc.). These waves will be considered in more detail in a subsequent
publication.

\vspace{0.3cm}
\noindent
{\it Acknowledgments}.
VU acknowledges support from the European Science Foundation (ESF) within
the framework of the ESF activity "The New Physics of Compact Stars".
VU thanks also the Russian Academy of Sciencs for financial support under the
programm OFN-15 and INAF-Ossevatorio Astrofisico di Catania for hospitality.

{}

\end{document}